\documentclass[twocolumn,showpacs,preprintnumbers,amsmath,amssymb,prl]{revtex4}

\usepackage{epsfig} 
\usepackage{graphicx}
\usepackage{dcolumn}
\usepackage{bm}

\begin{document}


\title{Delay Induced Instabilities in  Self-Propelling Swarms}

\author{Eric Forgoston}
 
\author{Ira B. Schwartz}
 
\affiliation{
Nonlinear Dynamical Systems Section, Plasma Physics Division, Code 6792, US Naval Research Laboratory, Washington, DC 20375, USA
}

\begin{abstract}
We consider a general model of self-propelling particles
interacting through a pairwise attractive force  in
the presence of noise and communication time delay.  Previous work by Erdmann,~{\it et al.}~[Phys. Rev. E {\bf 71}, 051904 (2005)] has shown
that a large enough noise intensity will cause a translating swarm of
individuals to transition to a rotating swarm with a stationary center of
mass. We show that with the
addition of a time delay, the model possesses a transition that depends on the
size of the coupling amplitude.  This transition is independent of the initial
swarm
state (traveling or rotating) and is characterized by the alignment of all of
the individuals along with a swarm oscillation.  By considering the mean field
equations without noise, we show that the time delay induced transition is
associated with a Hopf bifurcation.  The analytical result yields good
agreement with numerical computations of the value of the coupling parameter at the Hopf point. 
\end{abstract}


\maketitle

The collective motion of multi-agent systems have long been observed in
biological populations including bacterial colonies~\cite{bube95,b-jccvg97,blb98}, slime molds~\cite{lere91,nag98}, locusts~\cite{e-kwg98} and fish~\cite{ckjrf02}.  However, mathematical studies
of swarming behavior have been performed for only a few decades.  In addition
to providing examples of biological pattern formation, the information gained
from these mathematical investigations has led to an increased ability to intelligently design and control man-made vehicles~\cite{lefi01,jukr03,dcbc06,MorganS05,TriandafS05}.

Many types of mathematical models have been used to describe coherent swarms.
One popular approach is based on a continuum approximation
in which scalar and vector fields are used to describe all of the relevant
quantities~\cite{totu95,e-kwg98,totu98,fglo99,tobe04}.  Another popular approach is based on treating every biological or
mechanical individual as a discrete particle~\cite{vcb-jcs95,totu98,fglo99,ckjrf02,eem05}.
Depending on the problem, these individual-based models may be deterministic or stochastic.

Regardless of the type of swarm model being used, one can see the emergence of
ordered swarm states from an initial disordered state where individual
particles have random velocity directions~\cite{vcb-jcs95,totu95,totu98}.
These ordered states may be translational or rotational in motion, and they
may be spatially distributed or localized in clusters.  

In particular, it is known that a localized swarm
state may transition to a new dynamical region as the system parameters or the
noise intensity is changed.  For example, it has been shown
in~\cite{eem05} that a planar model of self-propelling particles interacting
via a harmonic attractive potential in the presence of noise  possesses a
noise-induced transition whereby the translational motion of the swarm breaks down
into rotational motion.

Another aspect of swarm modeling that has not yet been considered is the
effect of time delayed interactions arising from finite communication times
between individuals.  Much attention has been given to the effects of time
delays in the context of physiology~\cite{macgla77}, optics~\cite{ike79}, neurons~\cite{lansch07},
lasers~\cite{cskr06}, and many other types of systems.  The aim of this Letter is to
study the effect of a communication time delay on a model of self-propelling
individuals that interact through a pairwise attractive force in the presence
of noise.

We consider a general two-dimensional (2D) model of a swarm that consists of identical
self-propelled particles of unit mass.  The model is described by the
following evolution equations of motion:
\begin{equation}
\label{e:pos}
\dot{\bm r}_i={\bm v}_i,
\end{equation}
\begin{equation}
\label{e:vel}
\dot{{\bm v}}_i=(1-|{\bm v}_i|^2){\bm v}_i - {\bm V}_i + \xi_i(t),
\end{equation}
where ${\bm r}_i(t)$ and ${\bm v}_i(t)$ are respectively the 2D position and
velocity vectors of the $i^{\rm th}$ particle at time $t$.  The terms ${\bm v}_i$
and $-|{\bm v}_i|^2{\bm v}_i$ define respectively the mechanisms of
self-propulsion and frictional drag.
Therefore, if the last two terms on the right hand side of Eq.~(\ref{e:vel}) are neglected, the particles will
approach an equilibrium speed of $v_{eq}=1$.  

The term ${\bm V}_i$ in Eq.~(\ref{e:vel}) describes
the social interaction, or communication, of the $i^{\rm th}$ individual with all of the other
individuals.  There are many possible choices for ${\bm V}_i$ (e.g. Morse
function, power law function, etc.).  As an example, we define ${\bm V}_i$ as follows:
\begin{equation}
\label{e:V}
{\bm V}_i=\frac{a}{N}\mathop{\sum\limits_{j=1}^{N}}\limits_{i\ne j}({\bm r}_i(t) - {\bm r}_j(t-\tau)),
\end{equation}
where $a$ is the particle interaction coupling parameter, $N$ is the number of
particles, and $\tau$ is a constant communication time delay.  This particular
choice of ${\bm V}_i$ assumes that only pairwise interactions are important.
Furthermore, the interaction is purely attractive and grows linearly with the
separation between two particles, much like a spring potential.

Lastly, the term $\xi_i$ in Eq.~(\ref{e:vel}) describes a stochastic white
force of intensity $D$.  This noise is independent for different particles,
and is characterized by the following correlation functions:\\
$\ \langle\xi_i(t)\rangle =0, \ \ \langle\xi_i(t)\xi_j(t)\rangle =2D\delta(t-t')\delta_{ij}$.

We numerically integrate Eqs.~(\ref{e:pos})-(\ref{e:V}) using a stochastic
fourth-order Runge-Kutta scheme with a constant time step size of $0.001$.  To
achieve a traveling, localized swarm state, we used constant initial
conditions~\cite{note1} and switched on the noise after a short amount of time had passed.

It was shown
in~\cite{eem05} that the model described by Eqs.~(\ref{e:pos})-(\ref{e:V})
with $\tau=0$ (i.e. no time delay) possesses a noise-induced transition
whereby a large enough noise intensity causes a translating swarm of
individuals to transition to a rotating swarm with a stationary center of
mass, where the center of mass is defined as ${\bm R}(t)=(1/N)\sum_i{\bm r}_i(t)$. 
\begin{figure}[h]
\begin{minipage}{0.49\linewidth}
\includegraphics[width=4.3cm,height=3.0cm]{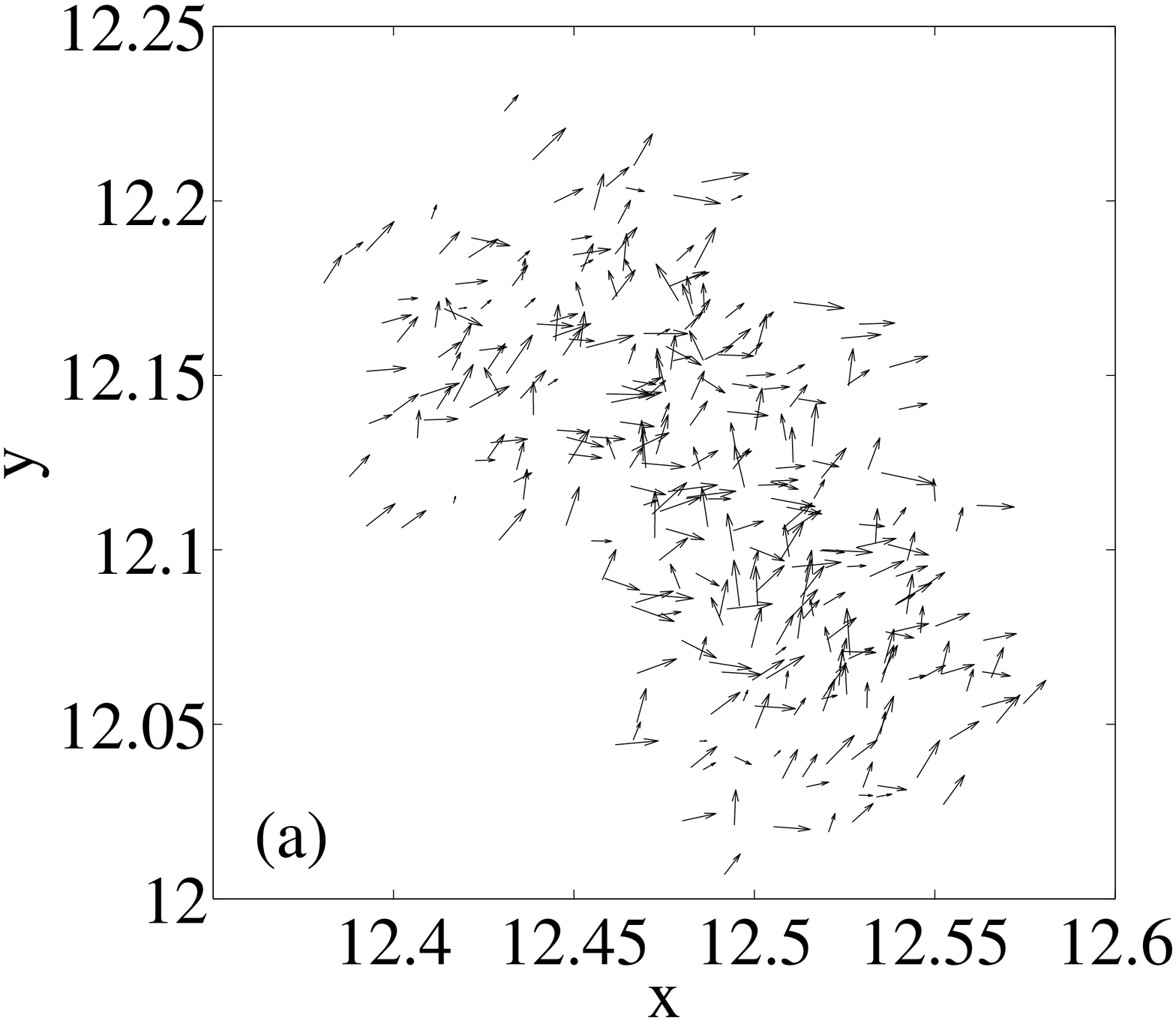}
\end{minipage}
\begin{minipage}{0.49\linewidth}
\includegraphics[width=4.3cm,height=3.0cm]{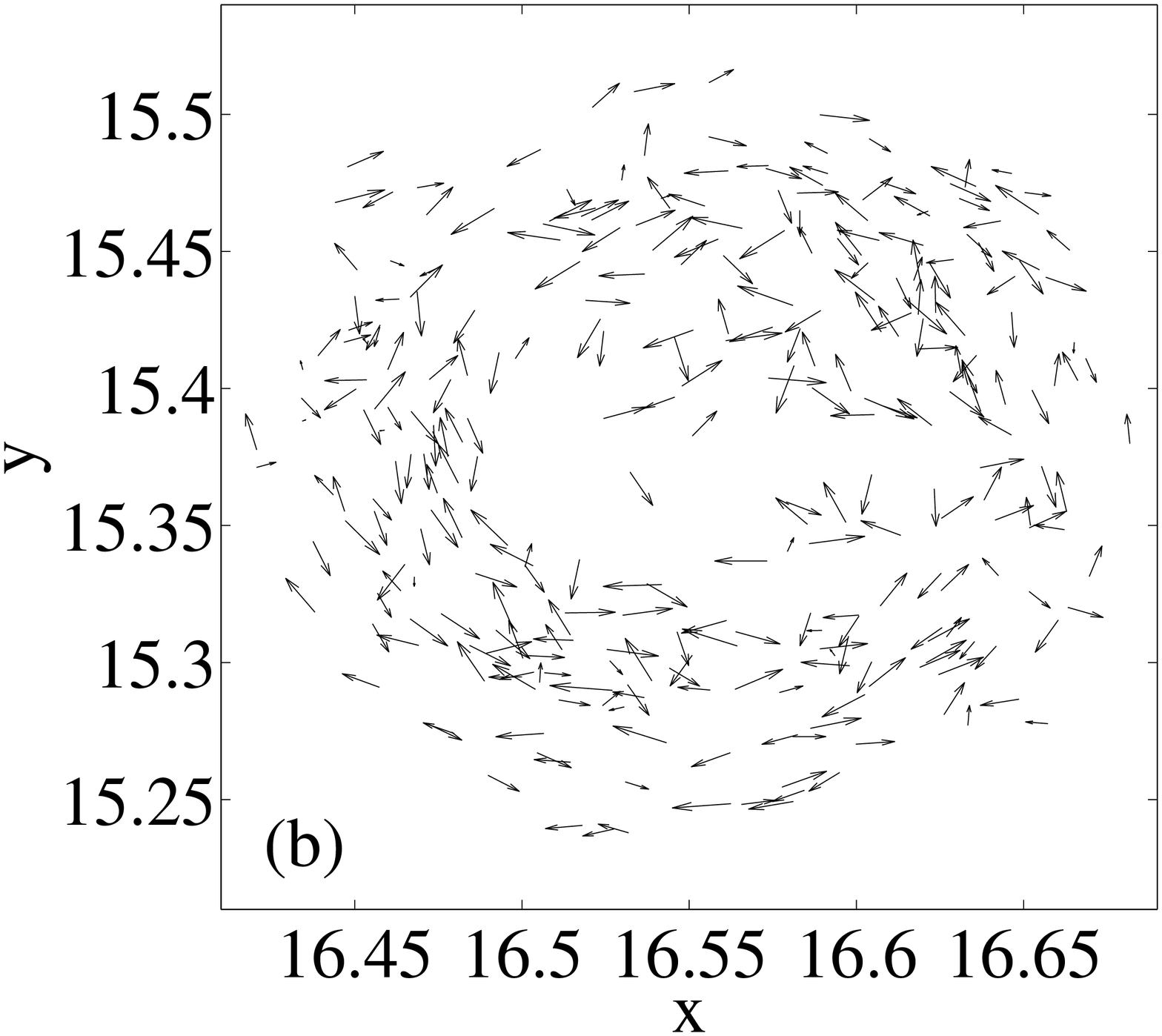}
\end{minipage}
\caption{\label{fig:no_delay_trans_rot} Snapshot of (a) a translating swarm
  (taken at $t=18$), and (b) a rotating swarm (taken at $t=40$), with $a=100$,
  $N=300$, $\tau = 0$, and $D=0.08$.}
\end{figure}

Figures~\ref{fig:no_delay_trans_rot}(a) and~\ref{fig:no_delay_trans_rot}(b) show
snapshots of a swarm at $t=18$ and $t=40$ respectively, with $a=100$, $N=300$,
$\tau=0$, and $D=0.08$.  The noise was switched on at $t=10$.  One can see that the
translating swarm [Fig.~\ref{fig:no_delay_trans_rot}(a)] has undergone a
noise-induced breakdown to become a rotating
swarm [Fig.~\ref{fig:no_delay_trans_rot}(b)].  For these values of $a$ and $N$, if a
noise intensity of $D<0.054$ is used,
then the swarm will continue to translate and it will not transition to a
rotational state~\cite{note2}.

Regardless of which state the swarm is in (translating or rotating), the
addition of a communication time delay leads to another type of transition.
This transition occurs if the coupling parameter, $a$, is large enough.  As an
example, we consider a swarm that has already undergone a noise-induced
transition to a rotational state before switching on the communication time
delay.  
\begin{figure}[h]
\begin{minipage}{0.49\linewidth}
\includegraphics[width=4.3cm,height=3.0cm]{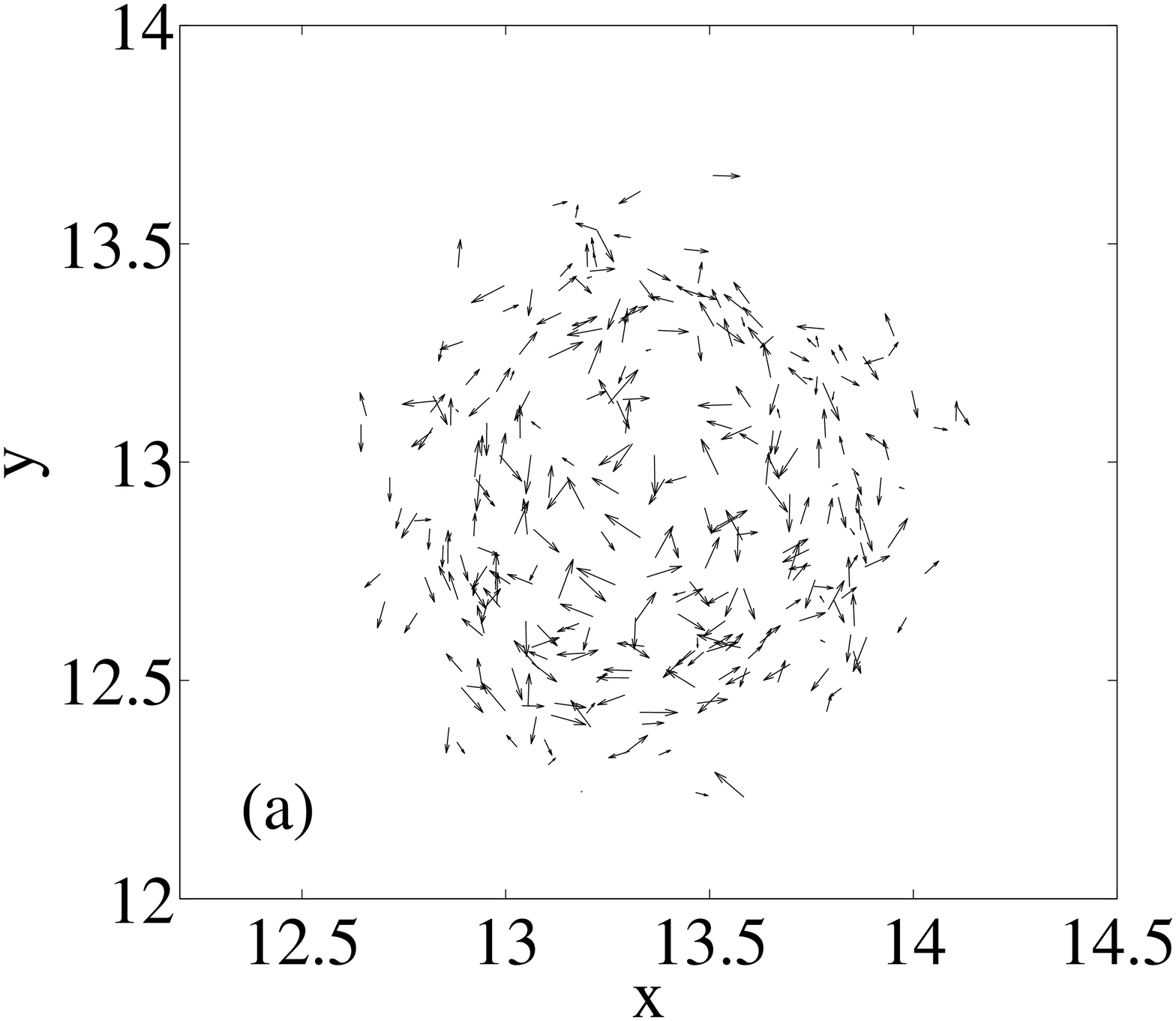}
\end{minipage}
\begin{minipage}{0.49\linewidth}
\includegraphics[width=4.3cm,height=3.0cm]{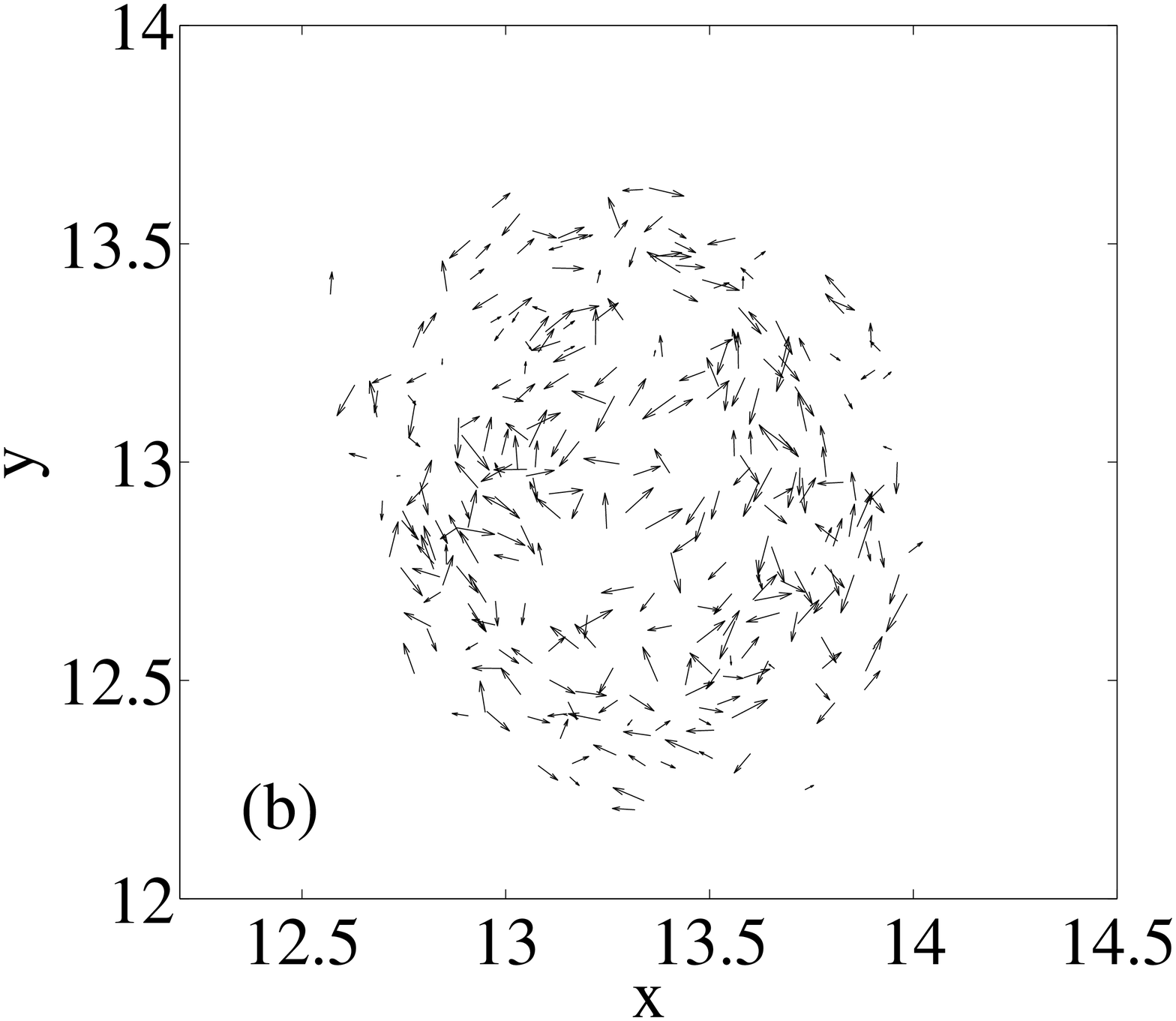}
\end{minipage}\\
\begin{minipage}{0.49\linewidth}
\includegraphics[width=4.3cm,height=3.0cm]{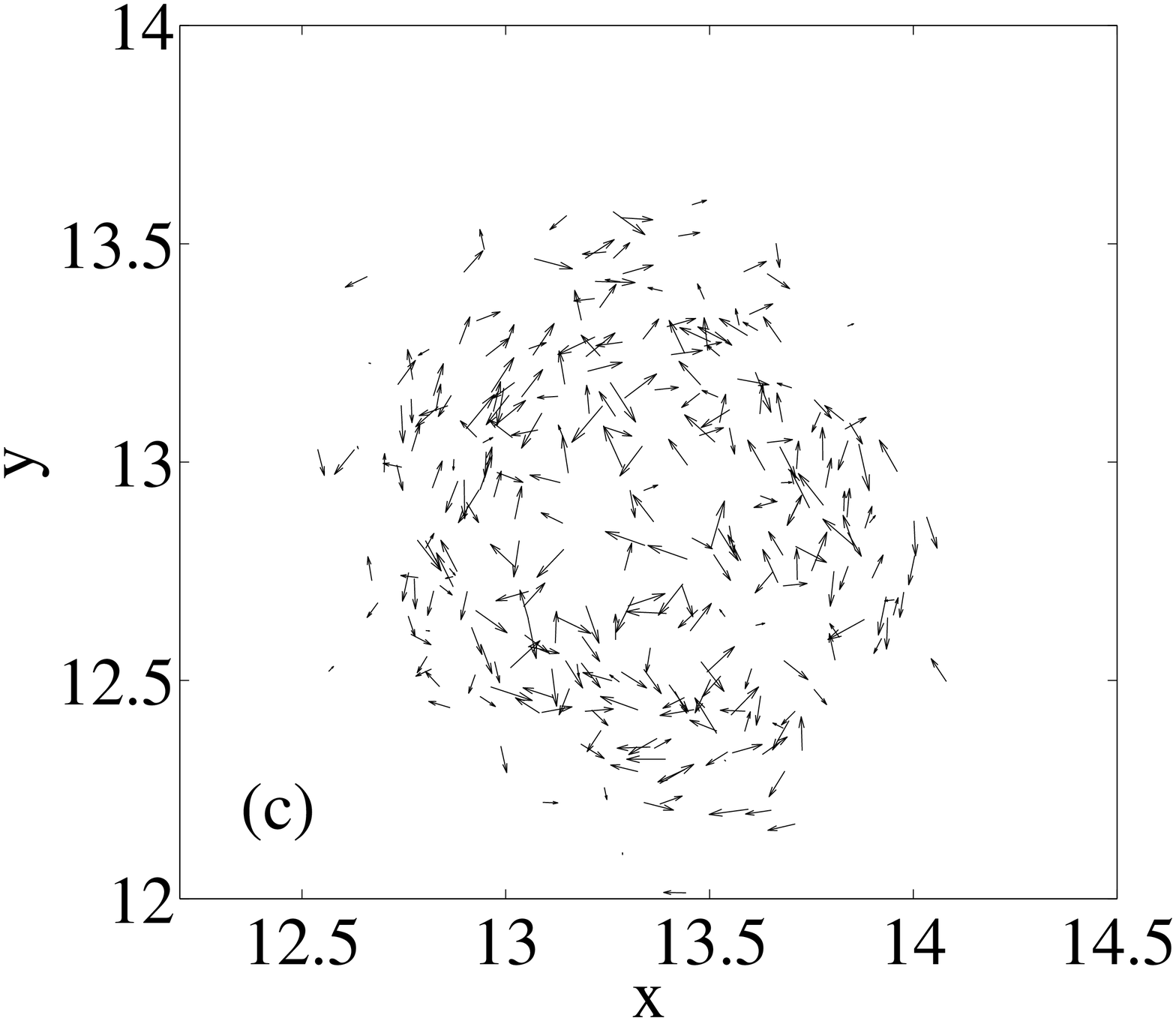}
\end{minipage}
\begin{minipage}{0.49\linewidth}
\includegraphics[width=4.3cm,height=3.0cm]{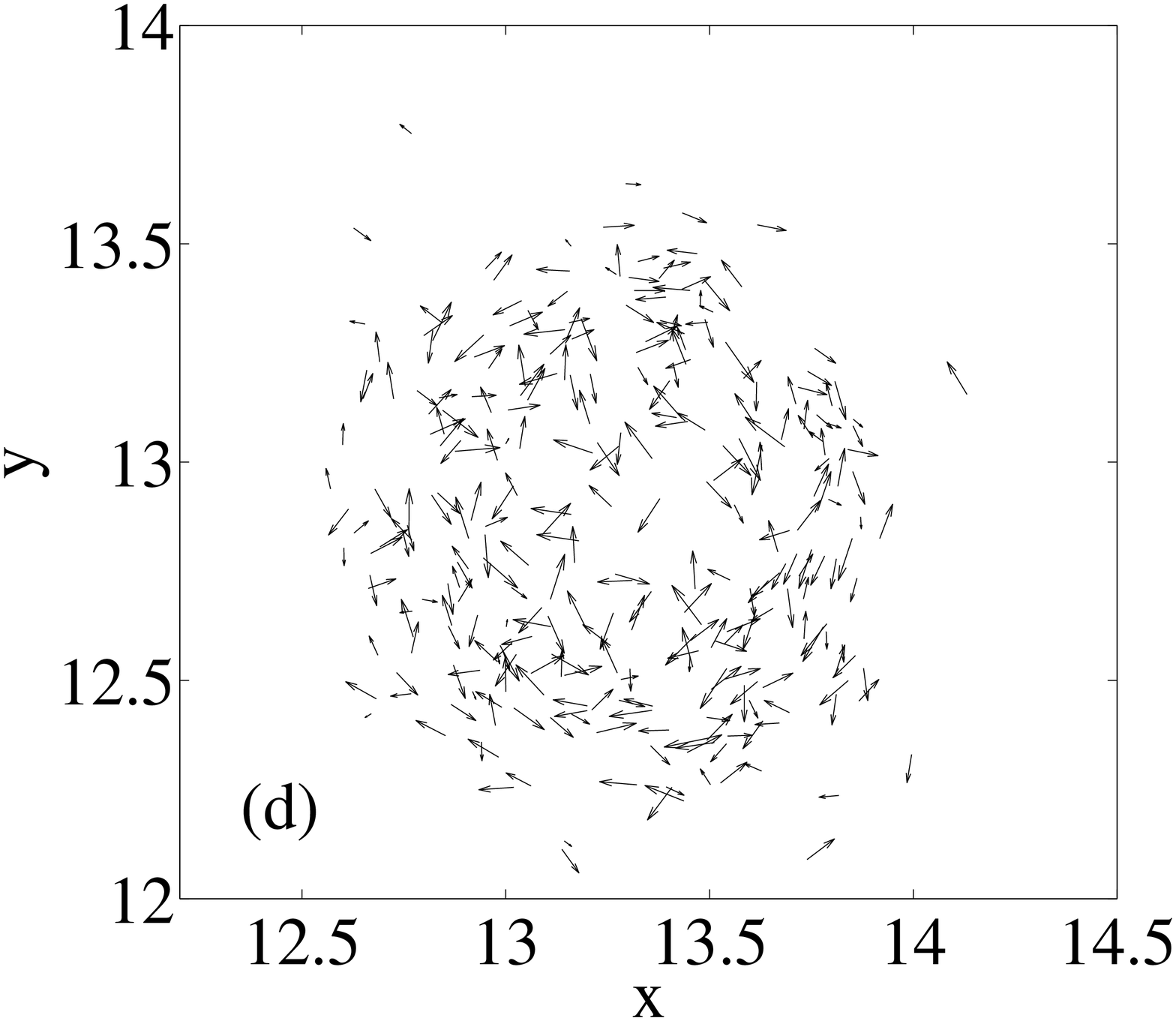}
\end{minipage}
\caption{\label{fig:delay_a2} Snapshots of a swarm taken at (a) $t=50$, (b)
  $t=100$, (c) $t=300$, and (d) $t=600$, with $a=2$, $N=300$, and $D=0.08$.
  The swarm was in a rotational state when the time delay of $\tau=1$ was switched on at $t=40$.}
\end{figure}

Figures~\ref{fig:delay_a2}(a)-\ref{fig:delay_a2}(d) show snapshots of a swarm at $t=50$,
$t=100$, $t=300$, and $t=600$ respectively, with $a=2$, $N=300$, $\tau=1$, and $D=0.08$.  The
noise was switched on at $t=10$, and since the noise intensity, $D$, is high
enough, the noise caused the swarm to transition to a rotating state [similar
to the one shown in Fig.~\ref{fig:no_delay_trans_rot}(b)].  With the swarm in
this stationary, rotating state, the communication time delay was switched on
at $t=40$.  One can see that for these values of time delay and coupling
parameter there is no qualitative change in the stationary, rotating swarm state.
\begin{figure}[h]
\begin{minipage}{0.49\linewidth}
\includegraphics[width=4.3cm,height=3.0cm]{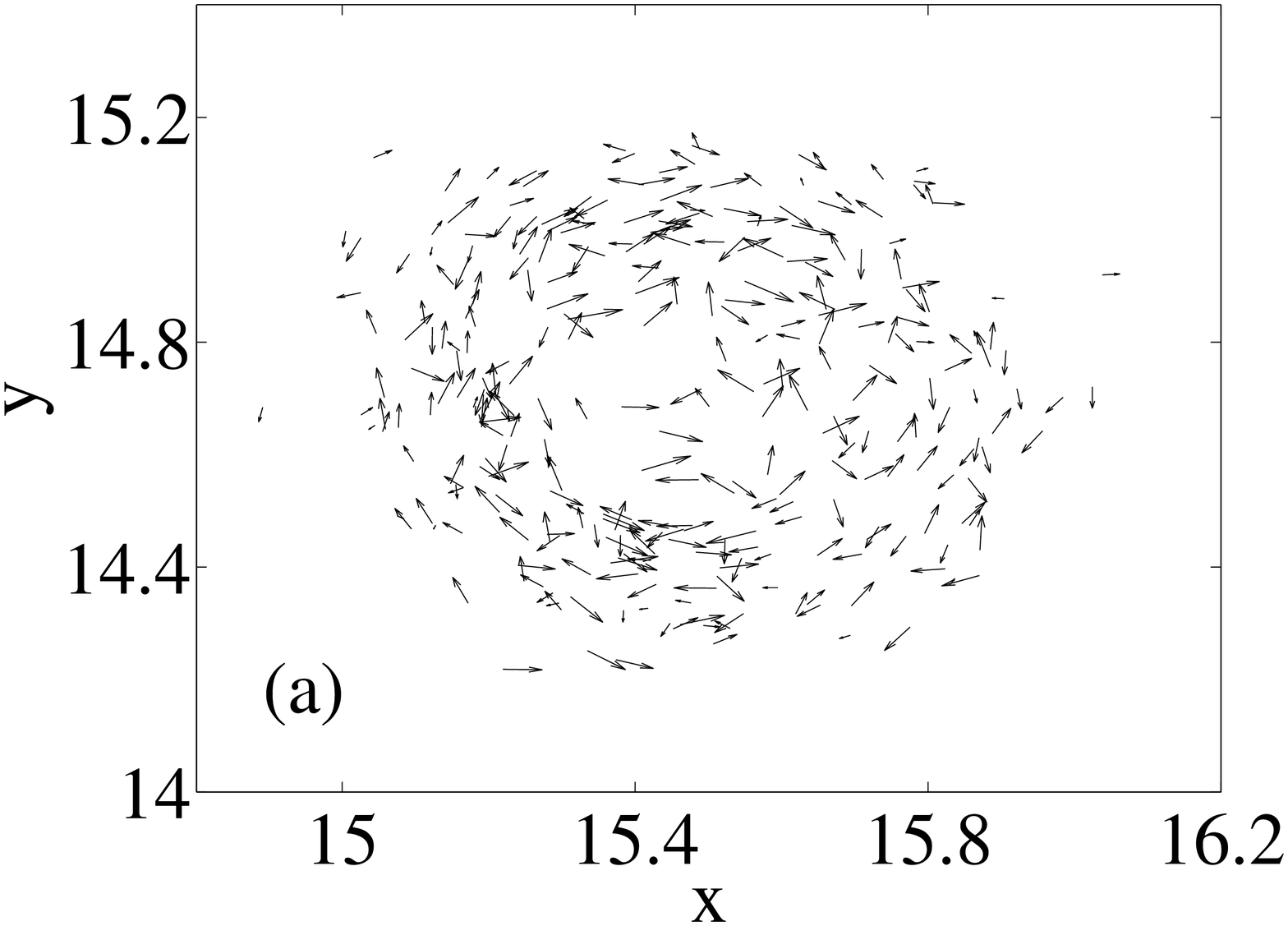}
\end{minipage}
\begin{minipage}{0.49\linewidth}
\includegraphics[width=4.3cm,height=3.0cm]{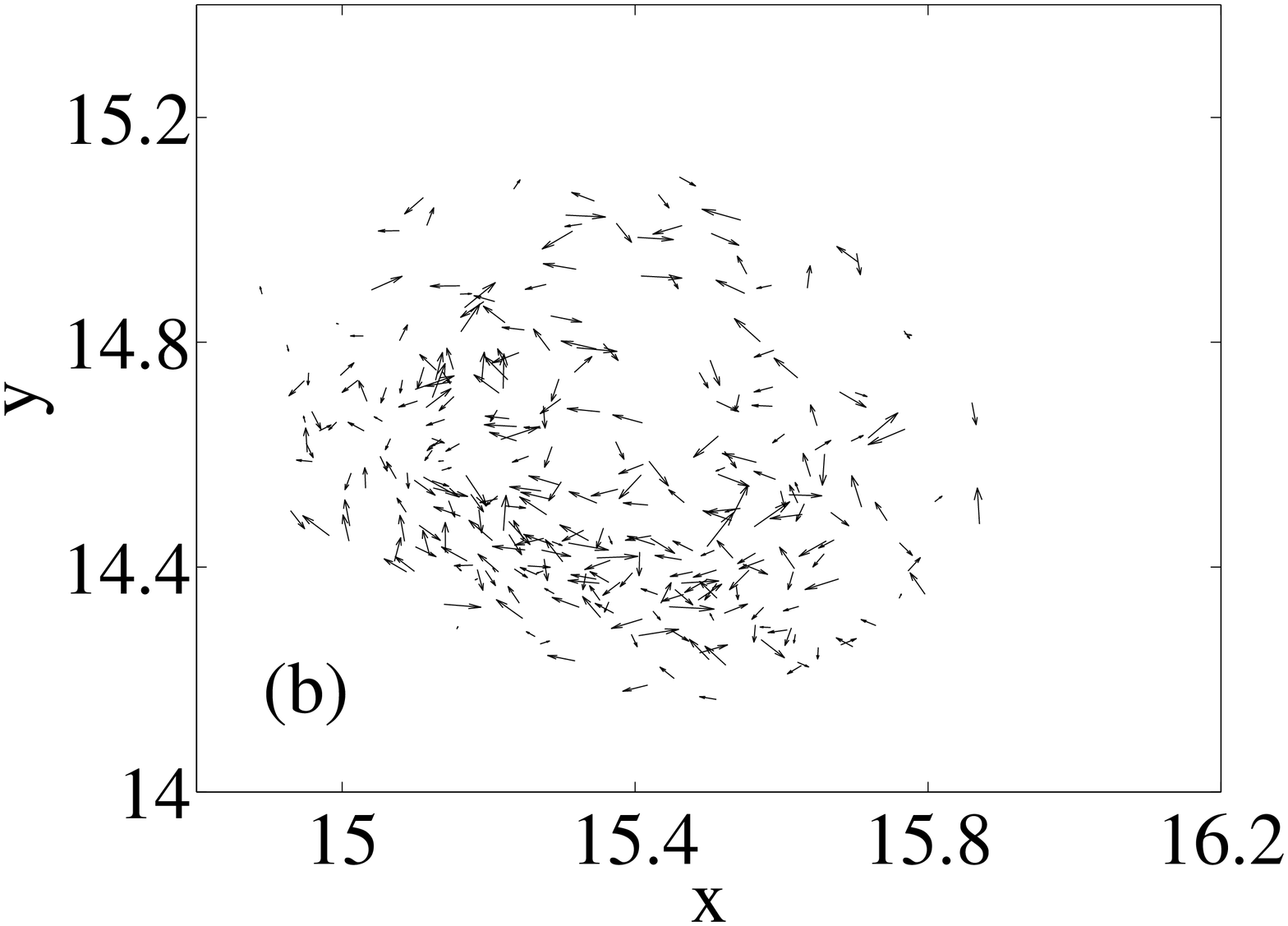}
\end{minipage}\\
\begin{minipage}{0.49\linewidth}
\includegraphics[width=4.3cm,height=3.0cm]{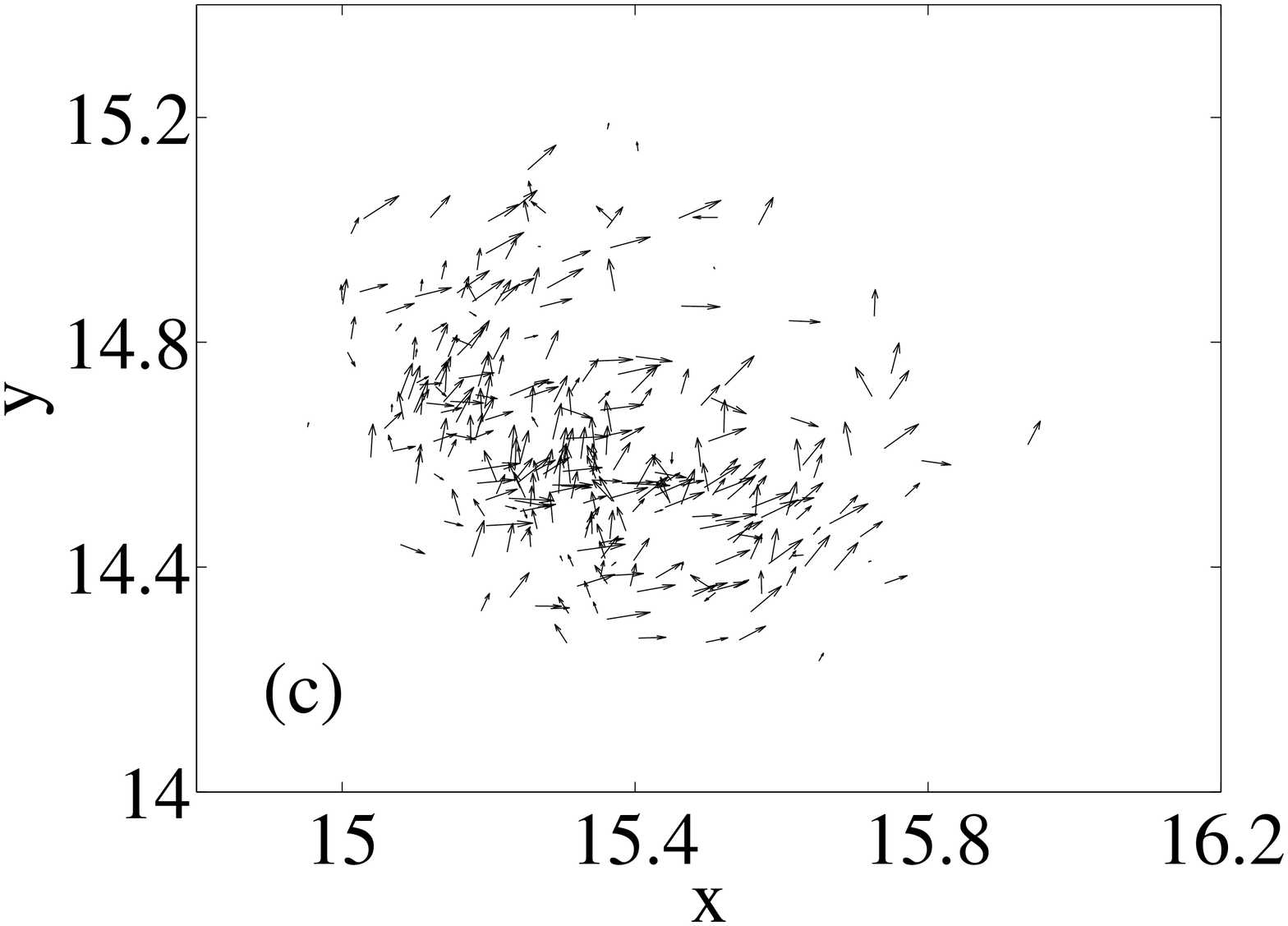}
\end{minipage}
\begin{minipage}{0.49\linewidth}
\includegraphics[width=4.3cm,height=3.0cm]{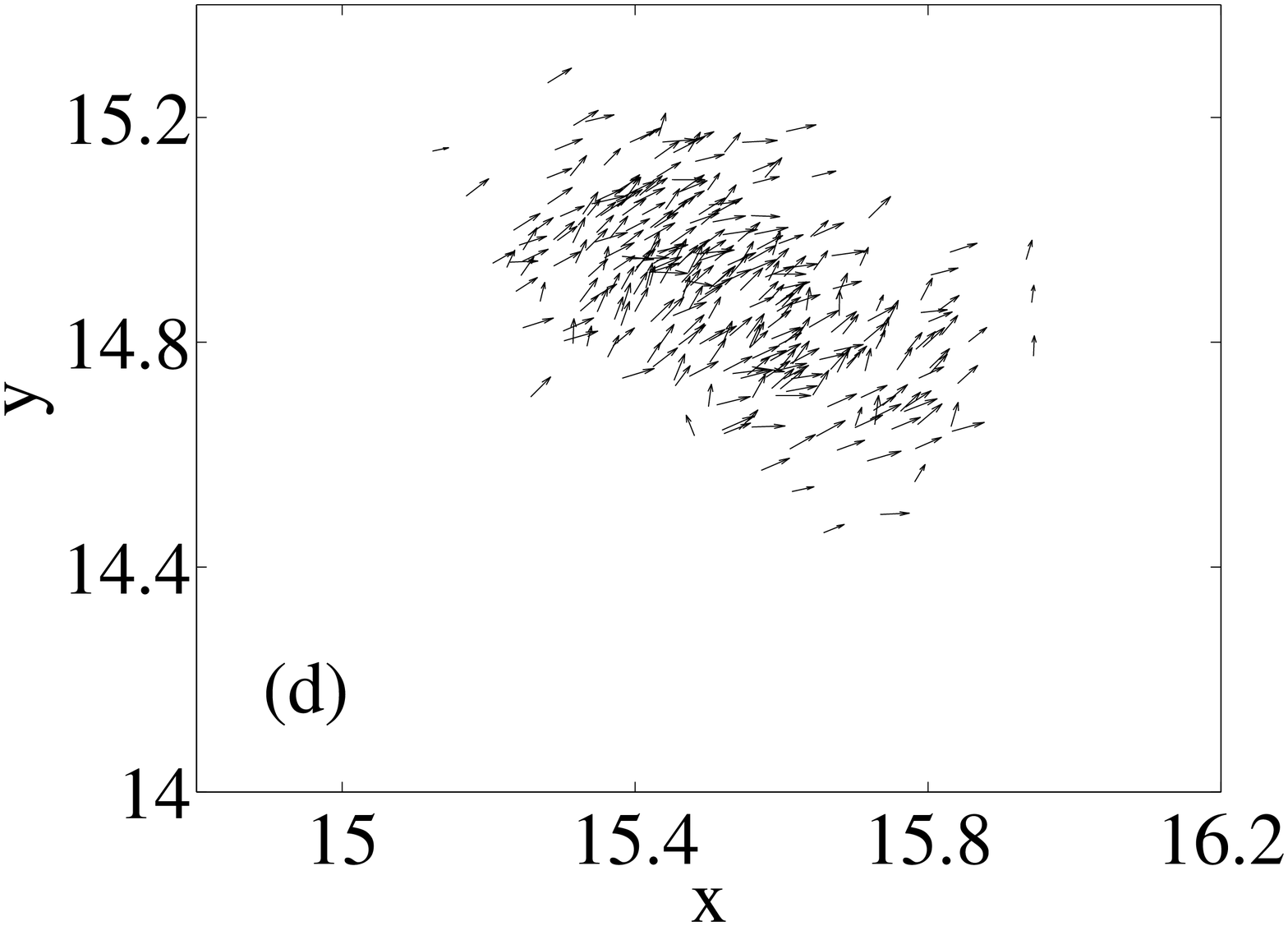}
\end{minipage}\\
\begin{minipage}{0.49\linewidth}
\includegraphics[width=4.3cm,height=3.0cm]{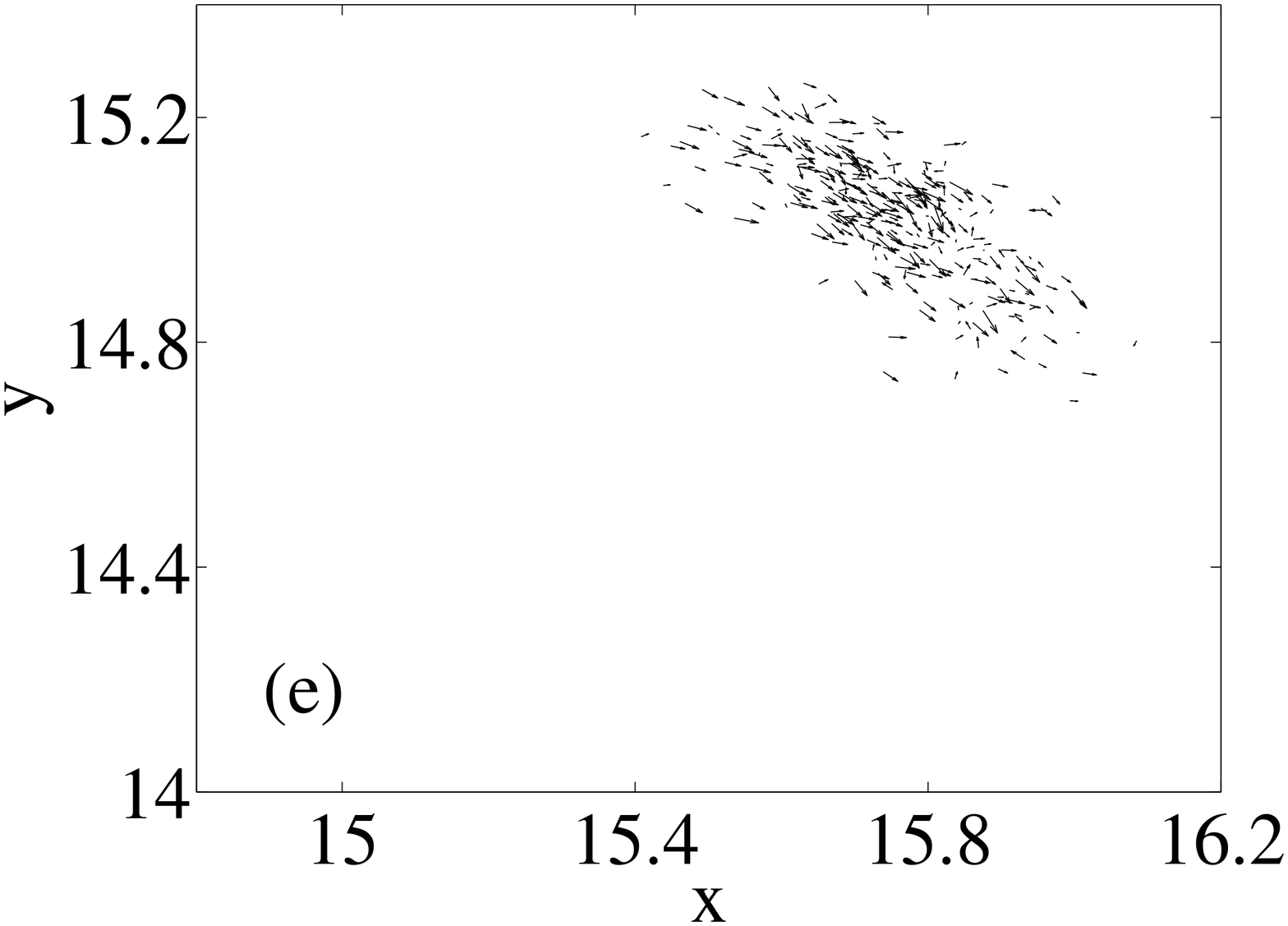}
\end{minipage}
\begin{minipage}{0.49\linewidth}
\includegraphics[width=4.3cm,height=3.0cm]{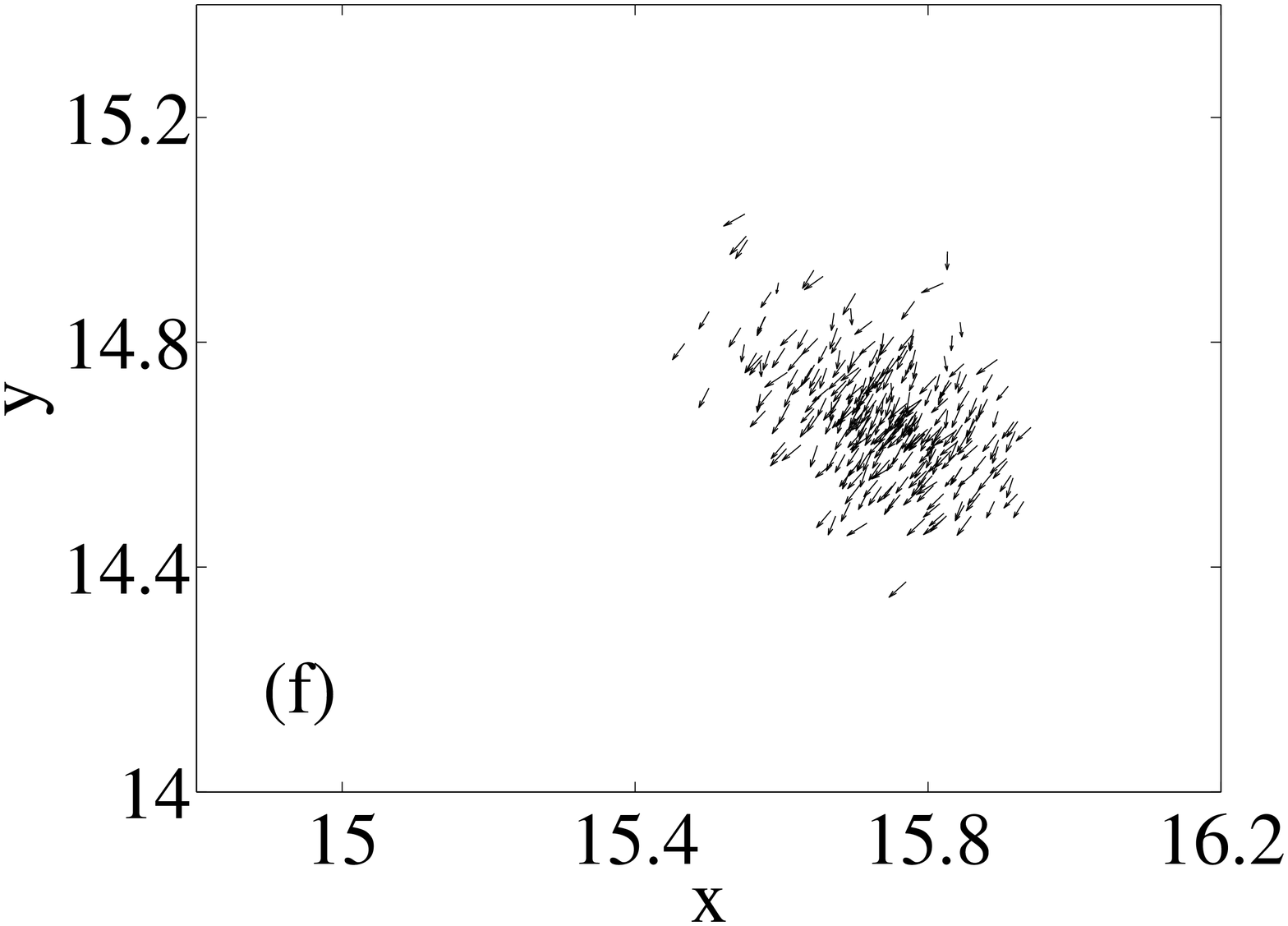}
\end{minipage}\\
\begin{minipage}{0.49\linewidth}
\includegraphics[width=4.3cm,height=3.0cm]{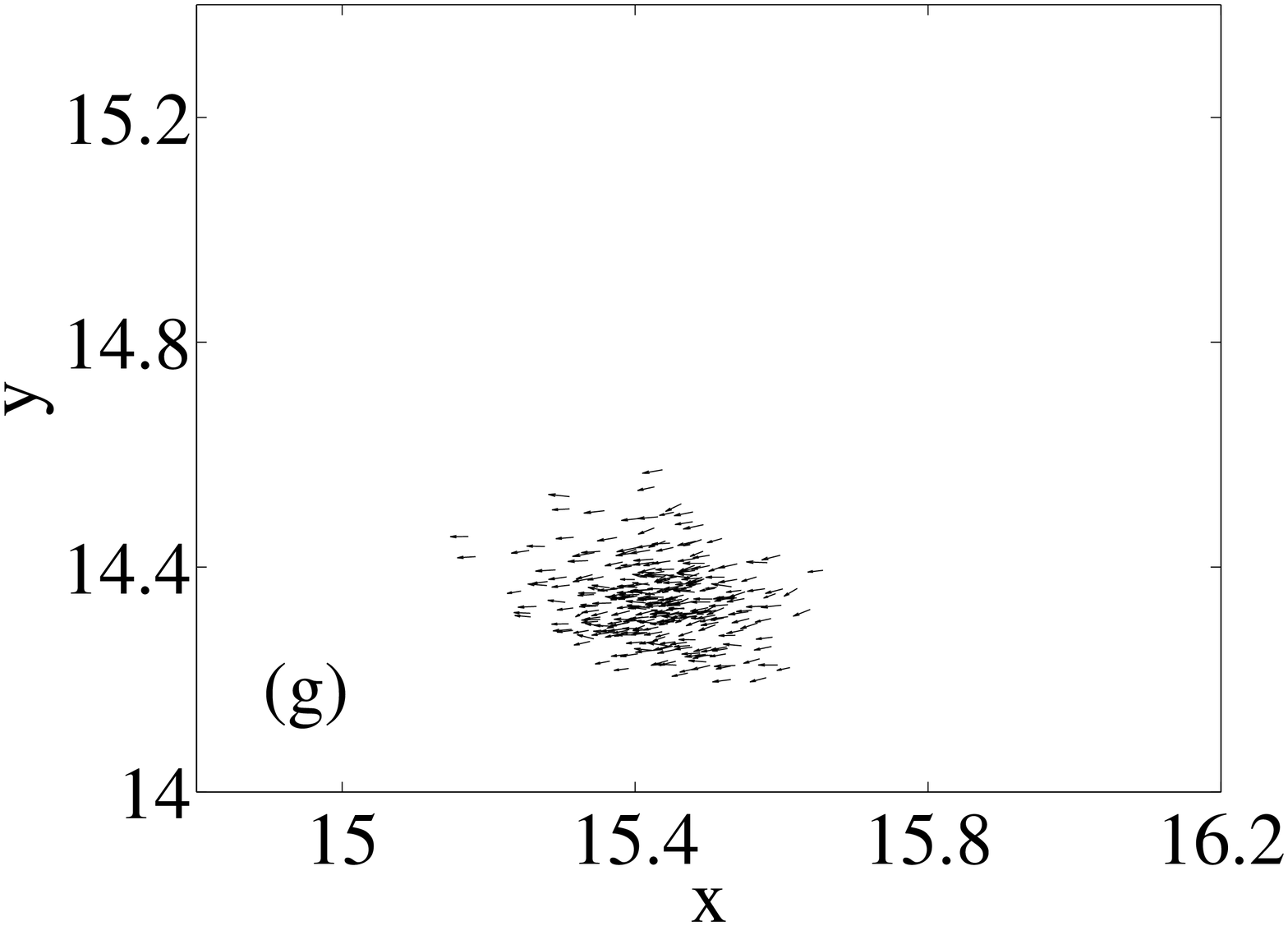}
\end{minipage}
\begin{minipage}{0.49\linewidth}
\includegraphics[width=4.3cm,height=3.0cm]{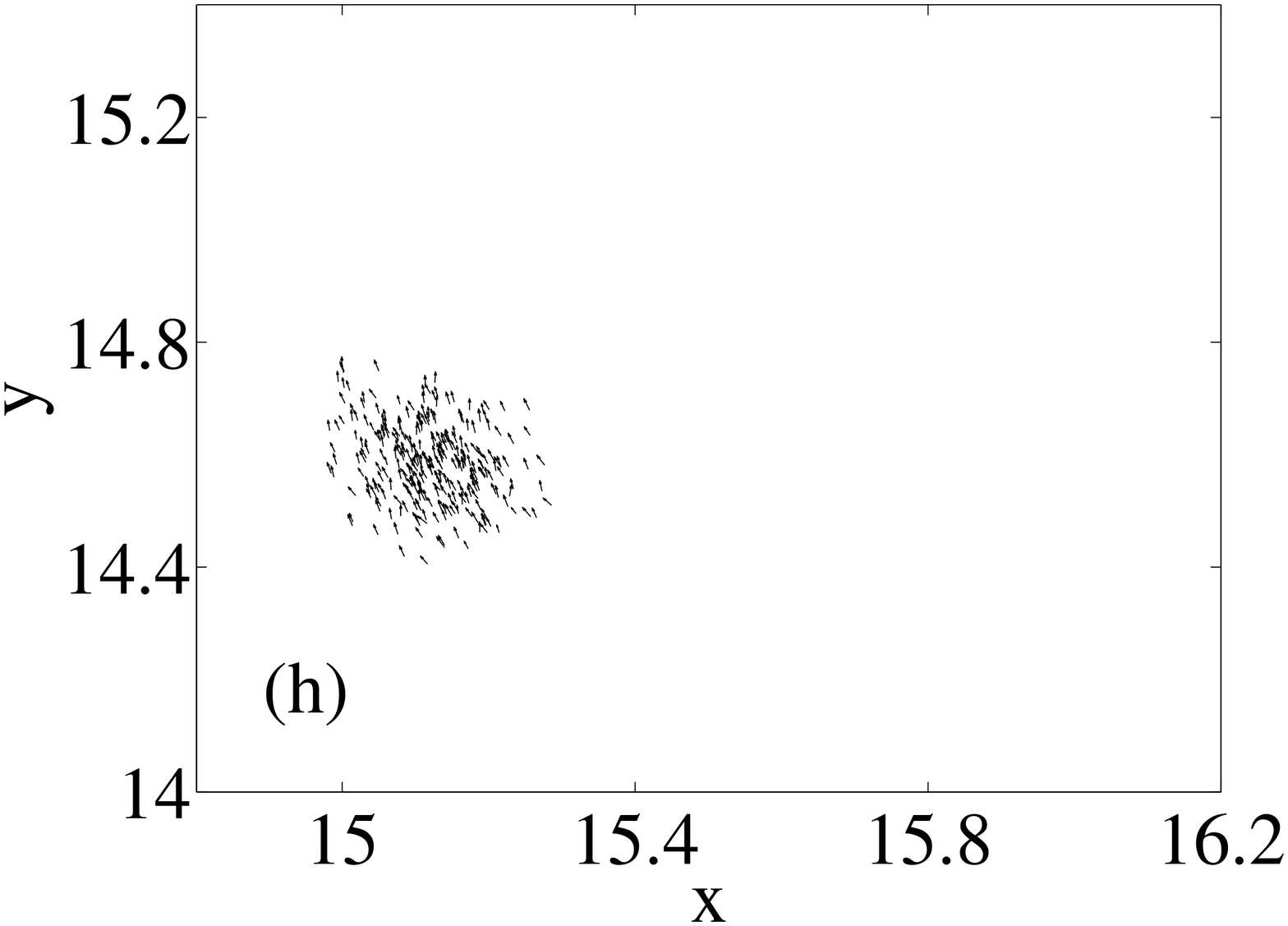}
\end{minipage}\\
\begin{minipage}{0.49\linewidth}
\includegraphics[width=4.3cm,height=3.0cm]{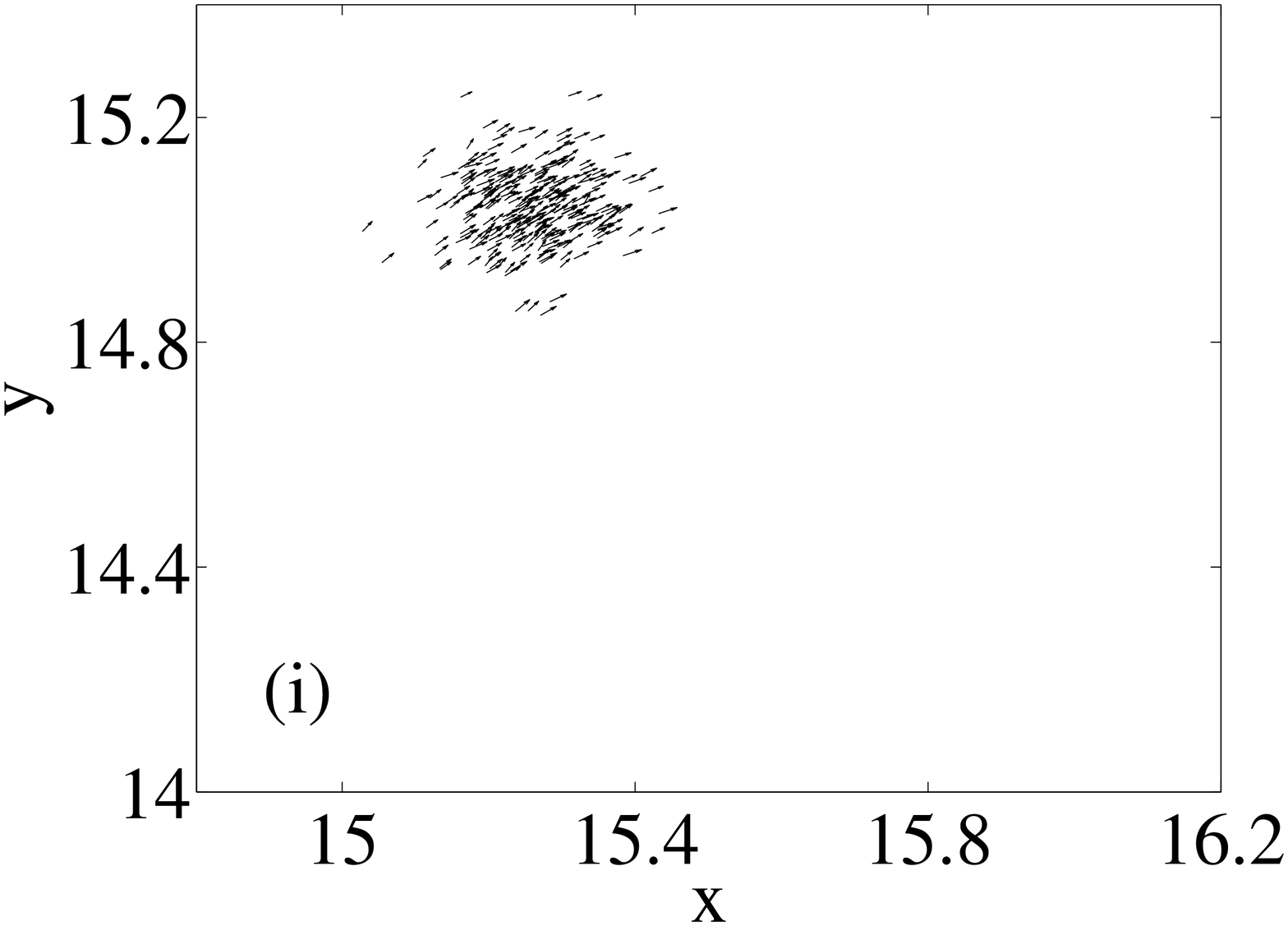}
\end{minipage}
\begin{minipage}{0.49\linewidth}
\includegraphics[width=4.3cm,height=3.0cm]{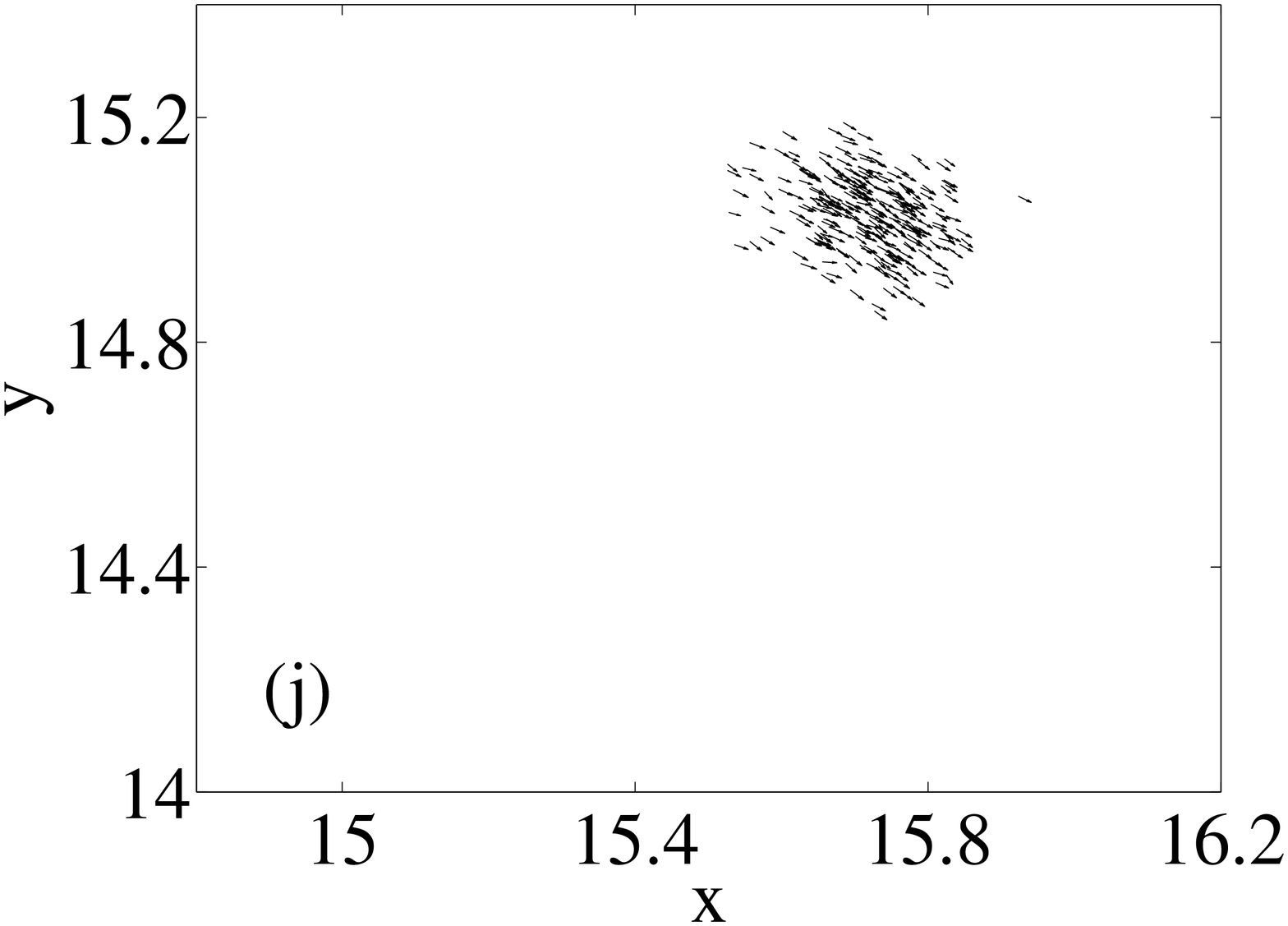}
\end{minipage}
\caption{\label{fig:delay_a4} Snapshots of a swarm taken at (a) $t=50$, (b) $t=60$, (c) $t=62$, (d) $t=64$, (e) $t=66$, (f) $t=68$, (g) $t=70$, (h) $t=72$, (i) $t=74$, and (j) $t=76$, with $a=4$, $N=300$, and $D=0.08$.  The swarm was in a rotational state when the time delay of $\tau=1$ was switched on at $t=40$.}
\end{figure}

\begin{figure}[h]
\includegraphics[width=8.6cm,height=6.0cm]{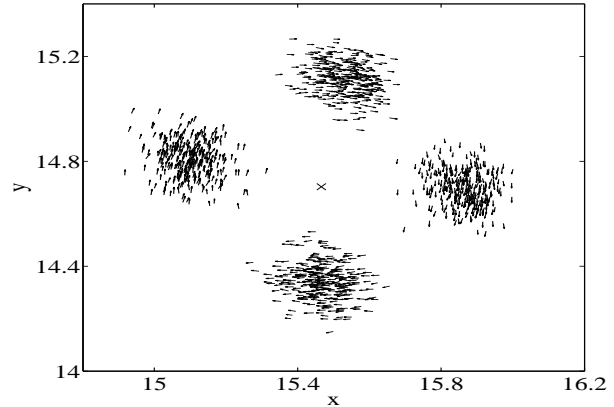}
\caption{\label{fig:oscillation} Motion of the oscillating swarm about the
  center of mass of the stationary, rotating swarm.  The oscillating swarm is
  shown at $t=90.2$ (left), $t=90.6$ (top), $t=91.0$ (right), and $t=91.4$ (bottom).  The location of the center of mass of the
swarm (at $t=40$) is denoted with a ``cross'' marker (center).}
\end{figure}

In contrast to this, Figs.~\ref{fig:delay_a4}(a)-\ref{fig:delay_a4}(j) show
snapshots of a swarm at $t=50$, $t=60$, $t=62$, $t=64$, $t=66$, $t=68$,
$t=70$, $t=72$, $t=74$, and $t=76$ respectively.  As in the previous case,
$N=300$, $\tau=1$, $D=0.08$, the noise was switched on at $t=10$ (causing the
swarm to transition to a stationary, rotating state), and once in this
rotating state, the time delay was switched on at $t=40$.  The only difference
is that now the value of the coupling parameter is $a=4$.  One can see that
with the evolution of time, the individual particles become aligned with one
another and the swarm becomes more compact.  Additionally, the swarm is no
longer stationary, but has begun to oscillate
[Figs.~\ref{fig:delay_a4}(g)-\ref{fig:delay_a4}(j)].  This clockwise oscillation can
more clearly be seen in Fig.~\ref{fig:oscillation}, which consists of the
center of mass, ${\bm R}$, of the
stationary, rotating swarm at $t=40$ (denoted by a ``cross'' marker) along
with snapshots of the oscillating swarm taken at
$t=90.2$, $t=90.6$, $t=91.0$, and $t=91.4$.  

This compact, oscillating aligned swarm state looks similar to
a single ``clump'' that is described in~\cite{dcbc06}.  However, where each
``clump'' of~\cite{dcbc06} contains only some of the total number of swarming
particles, our swarm contains every particle.  Additionally, while a
deterministic model along with global coupling is used to attain the
``clumps'' of~\cite{dcbc06}, our oscillating aligned swarm is attained with the use of
noise and a time delay. 

As we have shown, once the stochastic swarm is in the stationary, rotating state, the
addition of a time delay induces an instability.  At this point, the stochastic
perturbations have a minimal effect on the swarm.  Therefore, we will
investigate the stability of the stationary, rotating swarm state by deriving the mean field equations without noise.  The coordinates $x_i$ and
$y_i$ of each particle in the swarm can be written as follows:
\begin{equation}
\label{e:coord}
x_i=X+\delta x_i \ \ \ {\rm and} \ \ \ y_i=Y+\delta y_i,
\end{equation}
where $X$ and $Y$ are the coordinates of the center of mass, ${\bm R}$, of the
swarm.  Substitution of Eq.~(\ref{e:coord}) into the second-order differential
equation that is equivalent to Eqs.~(\ref{e:pos})-(\ref{e:V}) gives an
evolution equation for each $x_i$ and $y_i$.  Summing all $i$ of these
equations, using the fact that 
\begin{equation}
\label{e:mean}
\frac{1}{N}\sum\limits_{i=1}^{N}x_i(t)=X(t) \ \ \ {\rm and} \ \ \ \frac{1}{N}\sum\limits_{i=1}^{N}y_i(t)=Y(t),
\end{equation}  
and ignoring all fluctuation terms, leads to the following zero-order mean
field equations for the center of mass:
\begin{equation}
\label{e:X_mean_field}
\ddot{X}(t)=[(1-\dot{X}^2)\dot{X}-\dot{X}\dot{Y}^2](t)-a(X(t)-X(t-\tau)),
\end{equation}
\begin{equation}
\label{e:Y_mean_field}
\ddot{Y}(t)=[(1-\dot{Y}^2)\dot{Y}-\dot{Y}\dot{X}^2](t)-a(Y(t)-Y(t-\tau)).
\end{equation}

The steady state is given by $\dot{X}(t)=\dot{Y}(t)=0$, $X(t)=X(t-\tau)$, and
$Y(t)=Y(t-\tau)$.  Consideration of small disturbances about the steady state
allows one to determine the linear stability.  The
characteristic equation associated with the linearization of
Eqs.~(\ref{e:X_mean_field})-(\ref{e:Y_mean_field}) is
\begin{equation}
\label{e:char_eq}
\lambda(1-\lambda)+ae^{-\lambda\tau}-a=0,
\end{equation}
where the exponential term ${\rm exp}(-\lambda\tau)$ is due to the time delay
in the governing equations.  Since Eq.~(\ref{e:char_eq}) is transcendental
(which is often the case for delay differential equations),
there exists the possibility of an infinite number of solutions.  

Our numerical simulations indicate the existence of a supercritical Hopf bifurcation as the
value of the coupling parameter, $a$, is increased
(Figs.~\ref{fig:delay_a2}-\ref{fig:oscillation}).  We identify the Hopf
bifurcation point by choosing the eigenvalue to be purely imaginary.
Then our choice of $\lambda=i\omega$ is substituted into Eq.~(\ref{e:char_eq}).  The separation of
Eq.~(\ref{e:char_eq}) into real and imaginary parts leads to an equation for
the frequency, $\omega$, along with an equation for the value of $a$ at the
Hopf bifurcation point.  The two equations are
\begin{equation}
\label{e:freq}
\omega^2+\omega\cot{(\omega\tau)}-\omega\csc{(\omega\tau)}=0,
\end{equation}
\begin{equation}
\label{e:a}
a_H=\frac{\omega}{\sin{(\omega\tau)}}.
\end{equation}
Given a specific value of $\tau$, Eq.~(\ref{e:freq}) can be solved numerically
for $\omega$.  These values of $\tau$ and $\omega$ can then be substituted
into Eq.~(\ref{e:a}) to determine the value of $a$ at the Hopf point.
\begin{figure}[h]
\includegraphics[width=8.6cm,height=6.0cm]{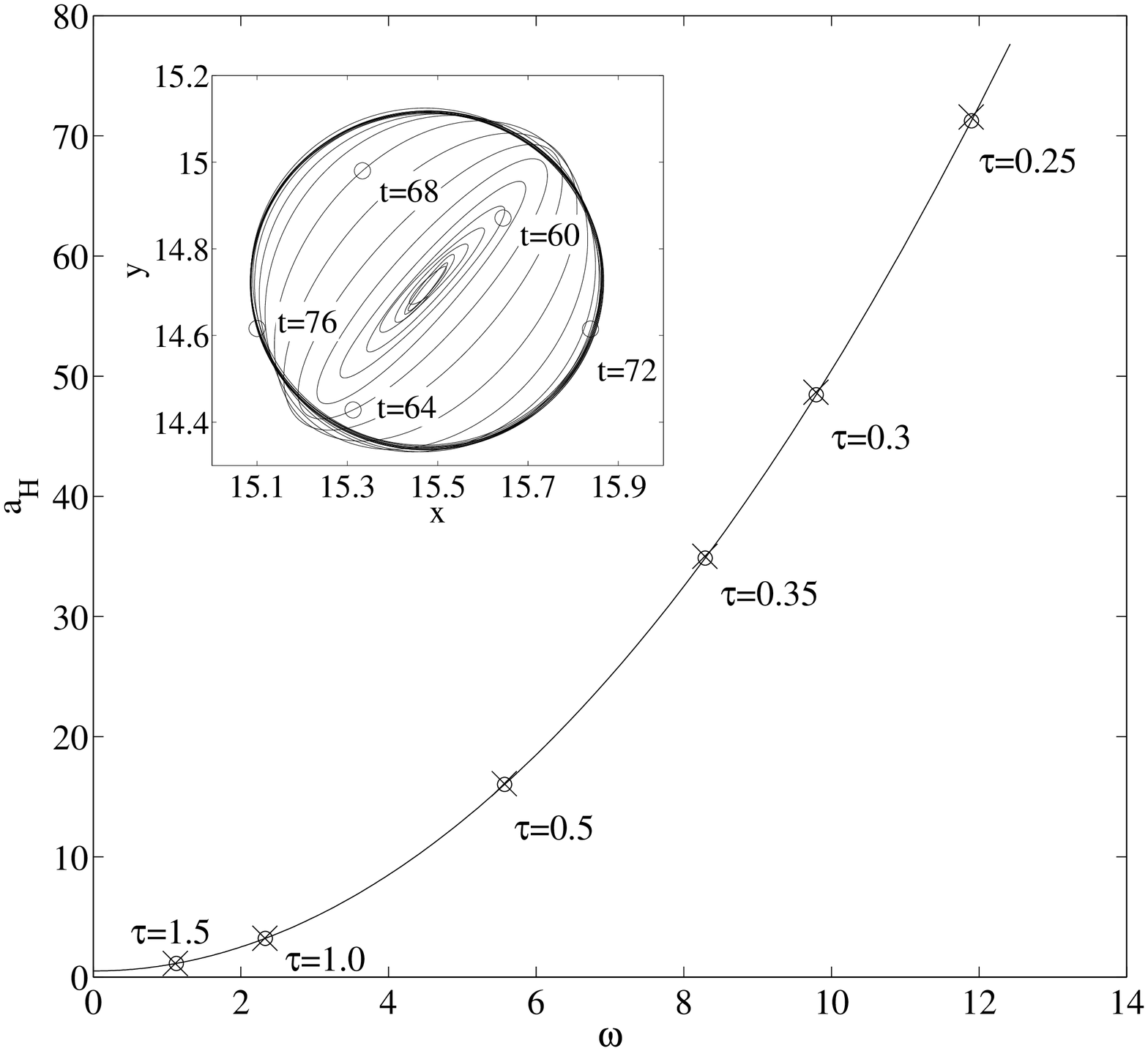}
\caption{\label{fig:Hopf}  Comparison of analytical (solid line) and numerical
(``cross'' markers) values of $a_H$ and $\omega$ for several choices of
$\tau$.  The analytical result is found using Eqs.~(\ref{e:freq})-(\ref{e:a}),
while the numerical result is found using a continuation
method~\cite{enlusa01} for Eqs.~(\ref{e:X_mean_field})-(\ref{e:Y_mean_field}).
The inset shows the stochastic trajectory of the center of mass of the swarm from $t=45$ to $t=90$ for the example shown in Fig.~\ref{fig:delay_a4}.}
\end{figure}

Figure~\ref{fig:Hopf} shows an excellent comparison of the analytical result given by
Eqs.~(\ref{e:freq})-(\ref{e:a}) with a numerical result which was found using
a continuation method~\cite{enlusa01}\nocite{enlusa01a} for
Eqs.~(\ref{e:X_mean_field})-(\ref{e:Y_mean_field}) for several choices of
$\tau$.  Furthermore, for $\tau=1$, the value of $a$ at the bifurcation point is
$a_H\approx 3.2$.  This value of $a_H$ corresponds very well to the change in
behavior of the stochastic swarm that was seen as the value of the coupling parameter was
increased from $a=2$ to $a=4$ (Figs.~\ref{fig:delay_a2}-\ref{fig:delay_a4}).

More evidence of the Hopf bifurcation is seen in the inset of
Fig.~\ref{fig:Hopf}.  The inset shows the stochastic trajectory of the center of mass of
the swarm from $t=45$ to $t=90$ for the example shown in Fig.~\ref{fig:delay_a4}.  Once the time delay is switched on at $t=40$
(with the swarm located at the center of the inset figure), the swarm begins
to oscillate.  The swarm moves along an elliptical path [the position of
its center of mass is denoted at several times that correspond to
Figs.~\ref{fig:delay_a4}(b),~\ref{fig:delay_a4}(d),~\ref{fig:delay_a4}(f),~\ref{fig:delay_a4}(h),
and~\ref{fig:delay_a4}(j)], until it eventually converges to the circular
limit cycle.

To summarize, we studied the dynamics of a self-propelling swarm in the
presence of noise and a constant communication time delay and prove that
the delay induces a transition that depends upon the size of the interaction
coupling coefficient.  Although our analytical and numerical results were
obtained using a model with linear, attractive interactions, the analysis may
be applied to models with more general forms of social interaction (these
results will appear elsewhere). 

Our results provide insight into the stability of complex systems comprised of
individuals interacting with one another with a finite time delay in a noisy
environment.  Furthermore, the results may prove to be useful in controlling
man-made vehicles where actuation and communication are delayed, as well as in
understanding swarm alignment in biological systems.  

We gratefully acknowledge support from the Office of Naval Research and the
Army Research Office.  E.F. is supported by a National Research Council
Research Associateship.  The authors benefitted from the valuable comments and
suggestions of anonymous reviewers.




\begin{thebibliography}{25}
\bibitem[{not({\natexlab{a}})}]{note1}
\bibinfo{note}{Although any constant value for the initial position and
  velocity may be used, we have arbitrarily chosen the following initial
  conditions: \ ${\bm r}_i={\bm 1}, \ {\bm v}_i={\bm 1}, \ i=1\ldots N$}.
\bibitem[{not({\natexlab{b}})}]{note2}
\bibinfo{note}{The critical noise intensity value reported in this article
  results from the use of a stochastic numerical solver that is of higher order
  than the one used in Ref.~\cite{eem05}, along with a Gaussian noise
  distribution.}
\end{thebibliography}

\end{document}